\documentclass[twocolumn,showpacs,preprintnumbers,amsmath,amssymb,prl]{revtex4}

\usepackage{graphicx}
\usepackage{dcolumn}
\usepackage{bm}
\usepackage{amssymb}
\begin{document}

\title{Collapse of electron band gaps in periodical structures dressed by a high-frequency field}
\author{O. V. Kibis}\email{Oleg.Kibis@nstu.ru}

\affiliation{Department of Applied and Theoretical Physics,
Novosibirsk State Technical University, Karl Marx Avenue 20,
630073 Novosibirsk, Russia}

%\date{\today}

\begin{abstract}
It is demonstrated theoretically that the strong interaction
between electrons in periodical structures and a high-frequency
electromagnetic field suppresses the Bragg reflection of the
electrons. As a result, the band gaps in electron energy spectra
of the structures (the Bragg gaps) are collapsed. This quantum
phenomenon can take place in various periodical structures,
including both natural crystalline solids and artificial
superlattices.
\end{abstract}


\maketitle

{\it Introduction.---} Advances in laser physics and microwave
techniques have made possible using a strong high-frequency
electromagnetic field  as a tool to manipulate electronic
properties of various quantum systems. Since the interaction
between electrons and the strong field cannot be described within
the conventional perturbation theory, the system ``electron +
field'' should be considered as a whole. Such a bound
electron-field object, which is known as an ``electron dressed by
field'' (dressed electron), became commonly used model in modern
physics \cite{Cohen-Tannoudji_b98,Scully_b01}. The field-induced
modification of energy spectrum of dressed electrons was studied
in both atomic systems \cite{Autler_55,Cohen-Tannoudji_b98} and
various condensed-matter structures, including bulk semiconductors
\cite{Elesin_69,Vu_04,Vu_05}, graphene
\cite{Lopez_08,Kibis_10,Glazov_14,Usaj_14}, quantum wells
\cite{Mysyrovich_86,Wagner_10,Kibis_12,Teich_13}, quantum rings
\cite{Kibis_11,Joibari_14,Kyriienko_15}, etc. In studies of
periodical structures dressed by a high-frequency field, the
research activity was focused substantially on the dynamic
localization of electrons
\cite{Dunlap_86,Holthaus_92,Holthaus_93,Platero_04}. As to the
effect of the field on electron band gaps in the structures, it
still awaited for detailed study. Performing theoretical analysis
of this open problem, I found that the field can collapse the
gaps.

It is well-known that the band gaps are the characteristic
property of electron energy spectra of all periodical structures.
Physically, they arise from the Bragg reflection of electron
waves, which was discovered at the dawn of quantum mechanics
(Nobel prize, 1937) and determines all electronic properties of
crystalline solids. It follows from the announced effect that a
high-frequency field suppresses the Bragg reflection of electrons.
As a consequence, periodical structures can be almost transparent
for electrons strongly coupled to the field. The present Letter is
devoted to the first theoretical analysis of this unexpected
quantum phenomenon which has direct relation to the basic
principles of condensed-matter physics.

{\it The Schr\"odinger problem.---} In order to simplify
calculations, let us consider a homogeneous dressing electric
field, $\mathbf{E}(t)=\mathbf{E}_0\sin\omega t$, where $E_0$ is
the amplitude of the field, and $\omega$ is the frequency of the
field. If an electron subjected to this dressing field is in
vacuum (or, generally, in a homogeneous medium), the Hamiltonian
of the dressed electron is
\begin{equation}\label{H0}
\hat{\cal{H}}_0=\frac{1}{2m}\left[\hat{\mathbf{p}}-\frac{e}{c}\mathbf{A}(t)\right]^2,
\end{equation}
where $\hat{\mathbf{p}}$ is the operator of electron momentum, $m$
is the electron mass, $e$ is the electron charge, and
$\mathbf{A}(t)=({c\mathbf{E}_0}/{\omega})\cos\omega t$ is the
vector potential of the considered dressing field. The accurate
solving of the Schr\"odinger equation with the Hamiltonian
(\ref{H0}),
$i\hbar{\partial\psi_\mathbf{k}(\mathbf{r},t)}/{\partial
t}=\hat{\cal{H}}_0\psi_\mathbf{k}(\mathbf{r},t)$, leads to the
exact wave function of the dressed electron \cite{Kibis_14},
\begin{eqnarray}\label{psi}
\psi_\mathbf{k}(\mathbf{r},t)&=&\exp\left[-i\left(\frac{\varepsilon_\mathbf{k}+\varepsilon_0}{\hbar}t
+\frac{E_0^2e^2}{8m\omega^3\hbar}\sin2\omega
t\right.\right.\nonumber\\
&-&\left.\left.\frac{e\mathbf{E}_0\mathbf{k}}{m\omega^2}\sin\omega
t\right)\right]\varphi_{\mathbf{k}}(\mathbf{r}),
\end{eqnarray}
where
$\varphi_{\mathbf{k}}(\mathbf{r})={V}^{-1/2}\exp(i\mathbf{k}\mathbf{r})$
is the plane electron wave, $\mathbf{k}$ is the electron wave
vector, $\mathbf{r}$ is the electron radius-vector, $V$ is the
normalization volume, $\varepsilon_\mathbf{k}={\hbar^2k^2}/{2m}$
is the energy spectrum of free electron, and
$\varepsilon_0={E_0^2e^2}/{4m\omega^2}$ is the field-induced shift
of the zero point of energy, which will be omitted in what
follows. It should be stressed that the Hamiltonian (\ref{H0})
with the same vector potential $\mathbf{A}(t)$ describes a
low-dimensional (two- or one-dimensional) electron system
subjected to a plane linearly polarized monochromatic
electromagnetic wave with the frequency $\omega$ and the amplitude
$E_0$, which propagates perpendicularly to the system. As a
consequence, the theory developed below is directly applicable,
particularly, to electrons in various nanostructures
(semiconductor quantum wires, carbon nanotubes, semiconductor
quantum wells, etc) irradiated by light.

Let an electron be in a periodical structure in the presence of
the same dressing field $\mathbf{A}(t)$. Then the wave function of
the electron, $\Psi(\mathbf{r},t)$, satisfies the Schr\"odinger
equation
\begin{equation}\label{SE}
i\hbar\frac{\partial\Psi(\mathbf{r},t)}{\partial t}=[\hat{\cal
H}_0+U(\mathbf{r})]\Psi(\mathbf{r},t),
\end{equation}
where $U(\mathbf{r})$ is the periodical potential of the
structure. Since the functions (\ref{psi}) with different wave
vectors $\mathbf{k}$ form the complete function system for any
time $t$, one can seek solution of the Schr\"odinger equation
(\ref{SE}) as an expansion
\begin{equation}\label{P}
\Psi(\mathbf{r},t)=\sum_{\mathbf{k}^\prime}a_{\mathbf{k}^\prime}(t)\psi_{\mathbf{k}^\prime}(\mathbf{r},t).
\end{equation}
Substituting the expansion (\ref{P}) into the Schr\"odinger
equation (\ref{SE}), we arrive at the expression
\begin{equation}\label{ak}
i\hbar\frac{\partial a_{\mathbf{k}}(t)}{\partial
t}=\sum_{\mathbf{k}^\prime}a_{\mathbf{k}^\prime}(t)
e^{i({\varepsilon}_\mathbf{k}-{\varepsilon}_{\mathbf{k}^\prime})
t/{\hbar}}e^{if_{\mathbf{k}^\prime-\mathbf{k}}\sin\omega
t}{U_{\mathbf{k}^\prime-\mathbf{k}}},
\end{equation}
where
$U_{\mathbf{k}}=({1}/{V})\int_VU(\mathbf{r})e^{i\mathbf{k}\mathbf{r}}\mathrm{d}^3\mathbf{r}$
is the Fourier transform of the periodical potential, and
$f_{\mathbf{k}}={e\mathbf{E}_0\mathbf{k}}/{m\omega^2}$.

It follows from the conventional Floquet theory of quantum systems
driven by an oscillating field
\cite{Zeldovich_67,Grifoni_98,Platero_04} that the wave function
(\ref{P}) has the form
$\Psi(\mathbf{r},t)=e^{-i\tilde{\varepsilon}(\mathbf{k})
t/\hbar}\phi(\mathbf{r},t)$, where the function
$\phi(\mathbf{r},t)$ periodically depends on time,
$\phi(\mathbf{r},t)=\phi(\mathbf{r},t+2\pi/\omega)$, and
$\tilde{\varepsilon}(\mathbf{k})$ is the quasi-energy of an
electron. It is well-known that the quasi-energy (the energy of
dressed electron) is the physical quantity which plays the same
role in quantum systems driven by an oscillating field as the
usual energy in stationary ones. Therefore, the present analysis
of the Schr\"odinger problem (\ref{SE}) is aimed to find the
energy spectrum, $\tilde{\varepsilon}(\mathbf{k})$. It follows
from the periodicity of the function $\phi(\mathbf{r},t)$ that one
can seek the coefficients $a_{\mathbf{k}}(t)$ in Eq.~(\ref{P}) as
a Fourier expansion,
\begin{equation}\label{aF}
a_{\mathbf{k}}(t)=e^{i[\varepsilon_\mathbf{k}-\tilde{\varepsilon}(\mathbf{k})]
t/\hbar}\sum_{n=-\infty}^{\infty}a_{n}({\mathbf{k}})e^{in\omega
t}.
\end{equation}
Substituting the expansion (\ref{aF}) into the expression
(\ref{ak}) and applying the Jacoby-Anger expansion,
$e^{iz\sin\theta}=\sum_{n=-\infty}^{\infty}J_n(z)e^{in\theta}$, to
transform the exponent in the right side of this expression, one
can rewrite the Schr\"odinger equation (\ref{ak}) as
\begin{equation}\label{Hk}
\sum_{n^\prime=-\infty}^{\infty}\sum_{\mathbf{g}}{\cal
H}_{nn^\prime}(\mathbf{g})a_{n^\prime}({\mathbf{k}+\mathbf{g}})=
\tilde{\varepsilon}(\mathbf{k})a_{n}({\mathbf{k}}),
\end{equation}
where
\begin{equation}\label{Hk0}
{\cal H}_{nn^\prime}(\mathbf{g})=
(\varepsilon_\mathbf{k}+n\hbar\omega)\delta_{n,n^\prime}\delta_{\mathbf{g},0}
+J_{n-n^\prime}\left(\frac{e\mathbf{E}_0\mathbf{g}}{m\omega^2}\right)U_{\mathbf{g}}
\end{equation}
is the Hamiltonian of dressed electron in the representation of
wave vectors, $J_n(z)$ is the Bessel function of the first kind,
$\mathbf{g}$ are the vectors of the reciprocal lattice of the
considered periodical structure, and $\delta_{m,n}$ is the
Kronecker symbol. It should be noted that the Schr\"odinger
equation (\ref{Hk}) describes still exactly the initial
Schr\"odinger problem (\ref{SE}). Next we will make some
approximations.

In what follows, let us assume that the field frequency, $\omega$,
is high to satisfy the condition
\begin{equation}\label{hf}
\left|\frac{{\cal
H}_{n0}(\mathbf{g})}{{\varepsilon_{\mathbf{k}}}-\varepsilon_{\mathbf{k}
+\mathbf{g}}+n\hbar\omega}\right|\ll1
\end{equation}
for $n\neq0$. Physically, the condition (\ref{hf}) means that the
field frequency lies far from resonant frequencies of the
periodical structure at the considered wave vector $\mathbf{k}$.
Mathematically, this makes it possible to treat the terms ${\cal
H}_{n0}(\mathbf{g})$ in the Hamiltonian (\ref{Hk0}) as a small
perturbation. Applying the conventional perturbation theory to
analyze the Schr\"odinger equation (\ref{Hk}) under the condition
(\ref{hf}), we easily arrive at the estimation
$|a_{n\neq0}({\mathbf{k}})|\ll1$. Since $a_{n}({\mathbf{k}})$ is
the quantum amplitude of the absorption (emission) of $n$ photons
by an electron, this estimation has the clear physical meaning:
The considered nonresonant field can be neither absorbed nor
emitted by the electron. As a consequence, the main contribution
to Eq.~(\ref{Hk}) arises from the terms with $n,n^\prime=0$, which
describe the elastic interaction between an electron and the
field. Neglecting the small terms with $n,n^\prime\neq0$,
Eq.~(\ref{Hk}) can be rewritten in the form
\begin{equation}\label{spec}
[\tilde{\varepsilon}(\mathbf{k})-\varepsilon_\mathbf{k}]a_0({\mathbf{k}})-
\sum_{\mathbf{g}}a_0({\mathbf{k}}+{\mathbf{g}})
\widetilde{U}_{\mathbf{g}}=0,
\end{equation}
where
\begin{equation}\label{U}
\widetilde{U}_{\mathbf{g}}=J_{0}\left(\frac{e\mathbf{E}_0\mathbf{g}}{m\omega^2}\right){U}_{\mathbf{g}}
\end{equation}
is the periodical potential renormalized by the high-frequency
field. Within the coordinate representation, the renormalized
potential (\ref{U}) reads as
\begin{equation}\label{RU}
\widetilde{U}(\mathbf{r})=\sum_{\mathbf{g}}J_{0}\left(\frac{e\mathbf{E}_0\mathbf{g}}{m\omega^2}\right){U}_{\mathbf{g}}e^{-i\mathbf{g}\mathbf{r}}.
\end{equation}
From the formal mathematical viewpoint, Eq.~(\ref{spec}) exactly
corresponds to the stationary Schr\"odinger problem with the
potential (\ref{RU}),
\begin{equation}\label{st}
\left[\frac{\hat{\mathbf{p}}^2}{2m}+\widetilde{U}(\mathbf{r})\right]\tilde{\psi}_\mathbf{k}(\mathbf{r})=\tilde{\varepsilon}(\mathbf{k})\tilde{\psi}_\mathbf{k}(\mathbf{r}).
\end{equation}
Indeed, expanding the stationary wave function,
$\tilde{\psi}_\mathbf{k}(\mathbf{r})$, on plane waves,
$\varphi_{\mathbf{k}}(\mathbf{r})$, and substituting this
expansion,
$\tilde{\psi}_\mathbf{k}(\mathbf{r})=\sum_\mathbf{k}a_0({\mathbf{k}})\varphi_{\mathbf{k}}(\mathbf{r})$,
into Eq.~(\ref{st}), we arrive at Eq.~(\ref{spec}). Thus, the
nonstationary Schr\"odinger problem (\ref{SE}) describing a
periodical structure dressed by a high-frequency nonresonant field
can be reduced to the conventional stationary Schr\"odinger
problem (\ref{st}) with the renormalized periodical potential
(\ref{RU}). This makes it possible to find the energy spectrum of
dressed electron in the periodical structure,
$\tilde{\varepsilon}(\mathbf{k})$, from the known energy spectrum
of bare electron in the same structure, $\varepsilon(\mathbf{k})$,
with the formal replacement
${U}_{\mathbf{g}}\rightarrow\widetilde{U}_{\mathbf{g}}$.

{\it The Bragg problem.---} The remarkable consequence of the
renormalization (\ref{U})--(\ref{RU}) is the field-induced
collapse of the electron band gaps which normally take place in
electron energy spectra of periodical structures at the borders of
the Brillouin zones. In order to demonstrate this effect, we will
restrict the following analysis by the simplest case of an
one-dimensional periodical potential $U(x)$ with the period $d$
but the proper generalization for any periodical structure can be
easily made. In the one-dimensional case, the band gaps take place
at the electron wave vectors $\mathbf{k}=\mathbf{g}_n/2$, where
$\mathbf{g}_n=(\pm2\pi n/d,\,0,\,0)$ are vectors of the reciprocal
lattice of the one-dimensional periodical structure, and
$n=1,2,3,...$ is the number of the corresponding Brillouin zone
(see Fig.~1a). Physically, the band gaps are originated in the
scattering of an electron by a periodical potential between the
two electron states with mutually opposite wave vectors,
$\mathbf{k}_\pm=\pm\mathbf{g}_n/2$ (the Bragg reflection of the
electron wave). Therefore, the main contribution to the band gaps
(the Bragg gaps) arises from the mixing of these two states by the
potential $U(x)$ \cite{Ziman_b72}. Within this conventional
approximation, the electron energy spectrum near borders of the
Brillouin zones is described by Eq.~(\ref{spec}), where all terms
should be omitted except the two terms corresponding to the
electron wave vectors $\mathbf{k}\approx\pm\mathbf{g}_n/2$. As a
result, Eq.~(\ref{spec}) turns into the two equations
\begin{eqnarray}\label{spec1}
[\tilde{\varepsilon}(\mathbf{k})-\varepsilon_\mathbf{k}]a_0({\mathbf{k}})-
\widetilde{U}_{\mathbf{g}_n}a_0({\mathbf{k}}+\mathbf{g}_n)&=&0,\nonumber\\
\left[\tilde{\varepsilon}(\mathbf{k})-\varepsilon_\mathbf{k}\right]
a_0({\mathbf{k}}+\mathbf{g}_n)-
\widetilde{U}_{-\mathbf{g}_n}a_0({\mathbf{k}})&=&0,
\end{eqnarray}
where the vectors $\mathbf{k}$ and $\mathbf{g}_n$ are assumed to
be mutually opposite directed. The solving of the two linear
algebraic equations (\ref{spec1}) results in the energy spectrum
of dressed electrons near the borders of the Brillouin zones,
\begin{equation}\label{EbB}
\tilde{\varepsilon}(\mathbf{k})=\frac{\varepsilon_\mathbf{k}
+\varepsilon_{\mathbf{k}+\mathbf{g}_n}}{2}\pm\frac{1}{2}\sqrt{(\varepsilon_\mathbf{k}
-\varepsilon_{\mathbf{k}+\mathbf{g}_n})^2+4|\widetilde{U}_{\mathbf{g}_n}|^2},
\end{equation}
which is plotted in Fig.~1a. Correspondingly, the Bragg gaps in
the energy spectrum (\ref{EbB}) read as
\begin{equation}\label{BG}
\Delta\tilde{\varepsilon}_n=2\left|J_{0}\left(\frac{e\mathbf{E}_0\mathbf{g}_n}{m\omega^2}\right){U}_{\mathbf{g}_n}\right|.
\end{equation}
The argument of the Bessel function in Eq.~(\ref{BG}) is the
parameter describing the interaction between an electron and a
dressing field. If the field is absent (${E}_0=0$), the gaps
(\ref{BG}) exactly coincide with the Bragg gaps in spectra of bare
electrons,
$\Delta{\varepsilon}_n=2\left|{U}_{\mathbf{g}_n}\right|$
\cite{Ziman_b72}. If the field is strong, the Bessel function
leads to the oscillating behavior of the gaps (\ref{BG}) and can
turn the gaps into zero (see Fig.~1b). Generally, the collapse of
the Bragg gaps in periodical structures corresponds to the zeros
of the renormalized periodical potential (\ref{U}), which are
defined by the condition
\begin{equation}\label{J}
J_{0}\left(\frac{e\mathbf{E}_0\mathbf{g}}{m\omega^2}\right)=0.
\end{equation}
Since the condition (\ref{J}) does not depend on concrete form of
the periodical potential, the discussed effect is of universal
physical nature and can take place in various periodical
structures, including both natural crystalline solids and
artificial superlattices.
\begin{figure}[th]
\begin{center}
\includegraphics[width=0.48\textwidth]{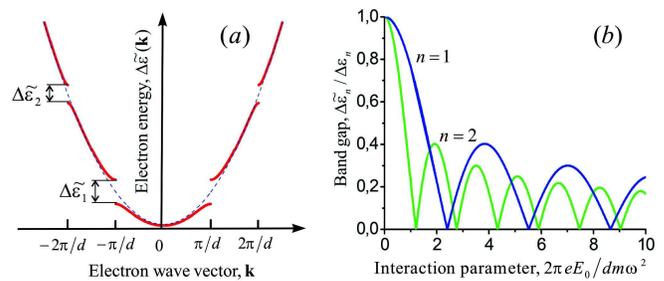}
\caption{(Color online) The band structure of an electron strongly
coupled to a high-frequency electromagnetic field in an
one-dimensional periodical potential with the period $d$: (a) The
energy spectrum of the electron in the scheme of extended zones
(solid line). The dashed line corresponds to the energy spectrum
of the free electron; (b) The dependence of the two first electron
band gaps on the parameter of the electron-field interaction.}
\end{center}
\end{figure}

From the physical point of view, the collapse of the Bragg gap
(\ref{BG}) means that a dressing field suppresses the Bragg
reflection of electron waves. In order to describe this phenomenon
accurately, let us assume that an electron is in the state
(\ref{psi}) with the wave vector $\mathbf{k}=\mathbf{g}_n/2$ at
the time $t=0$. Then the coefficients in the electron wave
function (\ref{P}) are
$a_{\mathbf{k}^\prime}(0)=\delta_{\mathbf{k}^\prime,\mathbf{k}}$.
Substituting the expansion (\ref{P}) into the Schr\"odinger
equation (\ref{SE}) and considering the periodical potential
$U(\mathbf{r})$ in the equation as a scattering perturbation, we
obtain the expression,
\begin{equation}\label{ak1}
a_{\mathbf{k}^\prime}(t)=-i\frac{U_{\mathbf{k}-\mathbf{k}^\prime}}{\hbar}
\int_0^te^{i(\varepsilon_{\mathbf{k}^\prime}-\varepsilon_\mathbf{k})t^\prime/{\hbar}}e^{if_{\mathbf{k}-\mathbf{k}^\prime}\sin\omega
t^\prime}\mathrm{d}t^\prime,
\end{equation}
which describes the amplitude of electron scattering from the
initial state $\mathbf{k}=\mathbf{g}_n/2$ to the final state
$\mathbf{k}^\prime$ in the first order of the conventional
perturbation theory. Applying the Jacoby-Anger expansion to
transform the exponent in the right side of Eq.~(\ref{ak1}), we
arrive from the scattering amplitude (\ref{ak1}) to the scattering
probability,
\begin{eqnarray}\label{wk}
|a_{\mathbf{k}^\prime}(t)|^2&=&\frac{\left|U_{\mathbf{k}^\prime\mathbf{k}}\right|^2}{\hbar^2}
\Bigg|\sum_{n=-\infty}^{\infty}{J_n\left(f_{\mathbf{k}-\mathbf{k}^\prime}\right)}\,
e^{i(\varepsilon_{\mathbf{k}^\prime}-\varepsilon_\mathbf{k}+n\hbar\omega)t/{2\hbar}}\nonumber\\
&\times&\int_{-t/2}^{t/2}e^{i(\varepsilon_{\mathbf{k}^\prime}-\varepsilon_\mathbf{k}
+n\hbar\omega)t^\prime/\hbar}\mathrm{d}t^\prime\Bigg|^2.
\end{eqnarray}
For long time $t\rightarrow\infty$, the integrals in the right
side of Eq.~(\ref{wk}) turn into the delta functions,
$\delta(\varepsilon_{\mathbf{k}^\prime}-\varepsilon_\mathbf{k}+n\hbar\omega)$.
Therefore, the scattering probability (\ref{wk}) reads as
\begin{eqnarray}\label{wk1}
|a_{\mathbf{k}^\prime}(t)|^2&=&4\pi^2\left|U_{\mathbf{k}-\mathbf{k}^\prime}\right|^2
\sum_{n=-\infty}^{\infty}J_n^2\left(f_{\mathbf{k}-\mathbf{k}^\prime}\right)\nonumber\\
&\times&\delta^2(\varepsilon_{\mathbf{k}^\prime}-\varepsilon_\mathbf{k}+n\hbar\omega).
\end{eqnarray}
In the considered case of nonresonant dressing field, the terms
with $n\neq0$ in Eq.~(\ref{wk1}) are turned into zero with the
delta functions. Transforming the remaining square delta function,
$\delta^2(\varepsilon_{\mathbf{k}^\prime}-\varepsilon_\mathbf{k})$,
with the conventional procedure,
$$\delta^2(\varepsilon)=\delta(\varepsilon)\delta(0)
=\frac{\delta(\varepsilon)}{2\pi\hbar}\lim_{t\rightarrow\infty}
\int_{-t/2}^{t/2}e^{i0\times
t^\prime/\hbar}dt^\prime=\frac{\delta(\varepsilon)t}{2\pi\hbar},$$
the probability (\ref{wk1}) can be rewritten as
\begin{equation}\label{W1}
w_{\mathbf{k}^\prime\mathbf{k}}=\frac{\mathrm{d}|a_{\mathbf{k}^\prime}(t)|^2}{\mathrm{d}
t}=\frac{2\pi}{\hbar}
\left|U_{\mathbf{\mathbf{g}}_n}\right|^2J^2_{0}\left(\frac{e\mathbf{E}_0\mathbf{g}_n}{m\omega^2}\right)
\delta(\varepsilon_{\mathbf{k}^\prime}-\varepsilon_\mathbf{k}).
\end{equation}
The nonzero delta function in Eq.~(\ref{W1}) corresponds
physically to the case of the  Bragg reflection,
$\mathbf{k}^\prime=-\mathbf{k}$. Therefore, Eq.~(\ref{W1})
describes the probability of the Bragg reflection per unit time.
As expected, the probability (\ref{W1}) turns into zero under the
condition (\ref{J}). This means that a periodical structure can be
almost transparent for electrons in the presence of a strong
high-frequency field.

{\it Discussion and conclusions.---} The expressions (\ref{BG})
and (\ref{W1}) are derived in the first-order (main) approximation
of the Bragg problem, which corresponds to the direct Bragg
scattering of electron waves between the two states,
$\mathbf{k}_\pm=\pm\mathbf{g}_n/2$. In many periodical structures,
this approximation is enough to describe electronic properties
adequately. However, the suppression of the direct Bragg
scattering requires to take into account the Bragg scattering
through intermediate electron states. Just this indirect
scattering process is responsible for the remanent Bragg gap when
the first-order gap (\ref{BG}) is zero. In order to estimate
values of the remanent gaps, let us consider a meander-like
one-dimensional periodical potential with the Fourier components
$U_{\mathbf{g}_n}=({V_0}/{n\pi})\sin\left({n\pi}/{2}\right)$,
which is commonly used to describe the potential relief of
semiconductor superlattices \cite{Ivchenko_b95}. Substituting the
Fourier components into Eq.~(\ref{spec}) and assuming the
condition (\ref{J}) to be satisfied, one can calculate the
remanent band gaps, $\delta\tilde{\varepsilon}_n$. In the case of
typical superlattice parameters $\Delta\varepsilon_n\sim10$~meV
and $d\sim10^{-9}$~m, the remanent gaps should be estimated as
negligibly small,
$\delta\tilde{\varepsilon}_n/\Delta\varepsilon_n\sim10^{-6}$.
Thus, a high-frequency field can collapse the Bragg gaps
effectively.

It should be noted that the Bragg reflection of electrons can be
suppressed in full for the special case of harmonic periodical
potential, $U(x)=V_0\cos(2\pi x/d)$, which can be realized,
particularly, in solids with using an acoustic wave
\cite{Keldysh_63}. Formally, this follows from the fact that the
harmonic potential has only nonzero Fourier component,
$U_{\mathbf{g}_1}=V_0/2$. Turning the sole renormalized Fourier
component, $\widetilde{U}_{\mathbf{g}_1}$, into zero with the
condition (\ref{J}), we turn the renormalized periodical potential
(\ref{RU}) into zero as a whole. Generally, one can conclude that
the renormalized periodical potential (\ref{RU}) can be
effectively controlled by a high-frequency field. This creates
physical prerequisites for the band engineering with the field.

In order to collapse the band gaps, the condition (\ref{hf})
should be satisfied for the electron wave vector at the border of
the Brillouin zone, $\mathbf{k}=\mathbf{g}/{2}$. Substituting the
energy spectrum (\ref{EbB}) into the inequality (\ref{hf}) and
keeping in mind that the unperturbed Bragg gap is
$\Delta{\varepsilon}\approx2\left|{U}_{\mathbf{g}}\right|$, one
can rewrite the condition (\ref{hf}) for this wave vector in the
well-behaved form, $\Delta\varepsilon/\hbar\omega\ll1$. Thus, we
have to satisfy both this high-frequency condition and the
gap-closing condition (\ref{J}). Since the condition (\ref{J}) can
be easily satisfied for low-frequencies $\omega$, periodical
structures with small band gaps are most appropriate from
experimental viewpoint. Particularly, the narrow-gap semiconductor
superlattices dressed by an infra-red irradiation seem to be most
promising. Let us assume that the field frequency lies in the
infra-red range, $\hbar\omega\sim10$~meV, the period of the
superlattice is $d\sim10^{-9}$~m, and the effective electron mass
in the semiconductor material is $m\sim10^{-29}$~g. Then the
condition (\ref{J}) can be satisfied for a dressing field with the
amplitude $E_0\sim10^3$~V/cm. Correspondingly, this relatively
weak dressing field can collapse the band gaps of meV scale. It
order to collapse greater band gaps, the irradiation intensity
should be increased. However, the increasing of stationary
irradiation can fluidize a condensed-matter periodical structure.
To avoid the fluidizing, it is reasonable to use narrow pulses of
irradiation which dress the periodical structure and narrow pulses
of a weak probing field which detect collapsing band gaps. This
pump-and-probe methodology has been elaborated long ago and is
commonly used to observe various dressing-field effects ---
particularly, modifications of energy spectrum of dressed
electrons arisen from the optical Stark effect
--- in semiconductor structures \cite{Joffre_88,Joffre_88_1,Lee_91,Vu_05}.
Since giant dressing fields (up to GW/cm$^2$) can be applied to
semiconductor structures within this approach, wide band gaps can
be collapsed with the pulsing fields.

Finalizing the discussion, it should be noted that the present
effect is conceptually opposite to the effect of dynamic
localization \cite{Dunlap_86,Holthaus_92,Holthaus_93,Platero_04}.
Formally, the both effects take place in periodical structures
driven by an alternating (ac) field. However, the first effect
collapses the electron band gaps, $\Delta\varepsilon$, and turns a
periodical structure to be transparent for electron waves, whereas
the second one collapses the allowed electron bands and leads to
the localization of electron wave packets. To avoid
misunderstandings, it should be stressed that these two different
effects take place in the two different frequency ranges of the
ac-field: The dynamic localization takes place at low frequencies
of the ac-field, $\omega$, which satisfy the condition
$\Delta\varepsilon/\hbar\omega\gg1$ \cite{Holthaus_92}, whereas
the present effect takes place at high-frequencies satisfying the
opposite condition $\Delta\varepsilon/\hbar\omega\ll1$. Thus, the
low-frequency and high-frequency ac-fields lead to the
substantially different electronic phenomena in periodical
structures.

Summarizing the aforesaid, one can conclude that a nonresonant
electromagnetic field can collapse band gaps in electron energy
spectra of periodical structures. Physically, the collapse of the
band gaps is caused by the field-induced suppression of the Bragg
reflection of electron waves. Therefore, periodical structures can
be almost transparent for electrons strongly coupled to the field.
Since the discussed effects are of universal nature, they can take
place in various periodical structures. Superlattices seem to be
most promising to observe these novel quantum phenomena.

{\it Acknowledgements.---} The work was partially supported by the
Russian Ministry of Education and Science, RFBR project No.
14-02-00033 and FP7 project QOCaN.

\end{document}